\documentclass[11pt]{article}
\usepackage{amsfonts}
\newtheorem{definition}{Definition}
\newtheorem{theorem}{Theorem}

\def\be{\begin{equation}}
\def\ee{\end{equation}}
\def\bea{\begin{eqnarray}}
\def\eea{\end{eqnarray}}

\def\>#1{{\bf #1}}

\def\1{\'{\i}}                           
\def\R{\rm I\kern-.2em R} 
\def\C{\rm I\kern-.5em C} 
\def\back{\!\!\!\!\!\!}

\def\tfrac#1#2{ {\scriptstyle { \frac {#1}{#2}}}}         
\def\pois#1#2{\left\{ {#1},{#2} \right\}}         
\def\conm#1#2{\left [ {#1},{#2} \right ]}

\def\Jt{J_3}
\def\Nt{N_3}
\def\Jpm{J_\pm}
\def\Npm{N_\pm}
\def\Nmp{N_\mp}
\def\Jp{J_+}
\def\Np{N_+}
\def\Jm{J_-}
\def\Nm{N_-}
\def\Ja{J_m}
\def\Na{N_m}

\def\luisH{{\cal H}}
\def\luisP{{\cal P}}
\def\luisM{{\cal M}}
\def\luisK{{\cal K}}
\def\luisD{{\cal D}}
\def\luisC{{\cal C}}

\begin{document}
\thispagestyle{empty}
\hfill\today 
\vspace{1.5cm}

\begin{center}

{\LARGE{\bf{COMODULE ALGEBRAS}}} 
\smallskip

{\LARGE{\bf{AND INTEGRABLE SYSTEMS}}} 
\smallskip

\vspace{1.4cm} 

{\large {\sc \'Angel Ballesteros}}\vspace{.2cm}\\
{\it Departamento de F\1sica, Universidad de Burgos\\
Pza. Misael Ba\~nuelos s.n., E-09001-Burgos, Spain} 
\vspace{.3cm}

{\large {\sc Fabio Musso}}\vspace{.2cm}\\
{\it SISSA\\
Via Beirut 2/4, Trieste, I-34013 Italy} 
\vspace{.3cm}

{\large {\sc Orlando Ragnisco}}\vspace{.2cm}\\
{\it Dipartimento di Fisica, Terza Universit\`a di Roma\\
Via della Vasca Navale 84, I-00146-Roma, Italy} 
\vspace{.3cm}

\end{center} 
  
\bigskip

\begin{abstract} 

A method to construct both classical and quantum completely integrable
systems from (Jordan-Lie) comodule algebras is introduced. Several 
integrable models based on a $so(2,1)$ comodule algebra, two
non-standard Schr\"odinger comodule algebras, the (classical and
quantum)
$q$-oscillator algebra and the Reflection Equation algebra are
explicitly
obtained.

\end{abstract} 

\medskip
\centerline{P.A.C.S.: 03.65 Fd, 02.20 Sv}
\vfill
\newpage

%%%%%%%%%%%%%%%%% INTRODUCTION %%%%%%%%%%%

\section{Introduction}

In this paper we present a generalization of the construction of
integrable systems with coalgebra symmetry given in \cite{BR} (let us
quote, for instance,
\cite{BHosc}-\cite{BRjmp} as different applications) by
using the notion of comodule algebras. Essentially,
the coalgebra approach \cite{BR} allowed the construction of integrable
systems on the space
$$\overbrace{A\otimes A\otimes\cdots\otimes A}^{N}$$
starting from a Hamiltonian
defined on a (either Poisson or non-commutative) Hopf algebra
$A$ with a number of casimir operators/functions. As we shall see in the
sequel, the comodule algebra approach will generalize this construction
to systems defined on 
$$\tilde A\otimes \overbrace{B\otimes\cdots\otimes B}^{N-1}$$
where
the ``one body'' Hamiltonian is again defined on the algebra $\tilde A$. 
But now $\tilde A$ has not to be a Hopf algebra, but only a
comodule algebra of another Hopf algebra $B$. We recall that
comodule algebras naturally appear, for instance, when the notion of
covariance is implemented in the context of noncommutative geometry (if
a given quantum space
$\cal{Q}$ is covariant under the action of a quantum group
$G_q$, the algebra
$\cal{Q}$ is a
$G_q$-comodule algebra \cite{Majid}-\cite{CD}). We would like to mention
that the existence of a Poisson analogue of such comodule algebra 
generalization was already pointed out in \cite{GMM}.

In Section 2 we introduce the general framework of Jordan-Lie
algebras \cite{Landsman} as a useful tool in order to describe
simultaneously the ``algebraic integrability" of both classical and
quantum systems.  Section 3 is devoted to the generalization of the
formalism presented in \cite{BR}. A first extension is obtained by
making use of homomorphisms of Jordan-Lie algebras with Casimir
elements. In this context, the comodule algebra construction
arises in a natural way when the previous homomorphism is just a
coaction between a given comodule algebra $A$ and a second Hopf
algebra
$B$. 

In order to illustrate the formalism, several integrable systems are
constructed in Section 4. The first one is derived from a coaction
$\phi:so(2,2)\rightarrow so(2,2)\otimes so(2,1)$. Secondly,
two new integrable deformations of the $N$-dimensional isotropic 
oscillator are constructed from two different (Poisson) Schr\"odinger
comodule
 algebras.
The
$q$-oscillator algebra \cite{Bi}-\cite{Kul} is also
shown to provide an example of integrable system with $su(2)_q$-comodule
algebra symmetry (this system was introduced for the first time in
\cite{Kul,ChaKul}). Moreover, a classical
$q$-oscillator is defined as a Poisson comodule algebra with respect
with the Poisson $su_q(2)$ algebra. From it, a classical
version of the Kulish Hamiltonian can be constructed, as well as the
classical analogue of the Jordan-Schwinger realization. 
 Finally, it
is shown how the Reflection Equation algebra \cite{KS} provides another
interesting example of $N$-dimensional integrable system related to
quantum spaces and endowed with comodule algebra symmetry (see also
\cite{Kul}). In the concluding Section we make a few comments
and outline some further generalizations of the method presented here.

\section{Jordan-Lie algebras and algebraic integrability}

\subsection{Jordan-Lie algebras}

We tersely recall the algebraic
structure of classical and quantum observables. A classical observable
is a smooth function
${\cal{F}}:{\cal P}
\rightarrow \mathbb{R}$  where ${\cal P}$ is a Poisson manifold.
The space of classical observables is naturally equipped with two
bilinear operations: the pointwise multiplication $\cdot$
\begin{displaymath}
(f \cdot g)(x):=f(x)g(x) \qquad x \in {\cal P}
\end{displaymath} 
and the Poisson bracket $[,]$ given in local coordinates by: 
\begin{displaymath}
[f,g](x):=\sum_{i,j}^N {\cal{P}}_{ij}(x) 
\frac{\partial f}{\partial x_i} \frac{\partial g}{\partial x_j}
\qquad x=(x_1,\dots,x_N)
\end{displaymath}
where ${\cal{P}}_{ij}$ is a Poisson tensor. The pointwise
multiplication is a unital, associative  and commutative 
product, while the Poisson bracket is a Lie product; the two structures
are coupled by
the Leibnitz rule (i.e. the Lie product is a derivation for the
associative product).

On the other hand, a quantum observable is a
self-adjoint operator on a  given Hilbert space. The space of quantum
observables is again equipped with two  bilinear operations defined
through the operator composition: the anticommutator
\begin{displaymath}
(a \cdot b):=ab+ba 
\end{displaymath}  
that is a commutative but not associative unital product, 
and the Lie product
\begin{displaymath}
[a,b]:=(ab-ba) 
\end{displaymath} 
We notice that both the ``classical" and the ``quantum" products obey
the Jordan identity, namely:
\begin{displaymath}
(a \cdot b) \cdot a^2= a \cdot (b \cdot a^2)
\end{displaymath}           
Hence both classical and quantum observables give rise to  
``Jordan-Lie algebras'' 
\cite{Landsman}. 

It is straightforward to generalize the definition of Jordan-Lie
algebra by replacing the Jordan product $\cdot$ with an associative but
not necessarily commutative product $\circ$. This is the kind of
Jordan-Lie algebra that we shall consider in this paper: while the
classical product will be again given by pointwise multiplication, the
quantum product will be just the operator (noncommutative)
composition:
\begin{displaymath}
(a \circ b):=a\,b
\end{displaymath}

\begin{definition}
A generalized Jordan-Lie algebra is a vector space ${\cal{A}}$
equipped with two bilinear maps $\circ$ and $[,]$ such that for all
$a,b,c \in {\cal{A}}$:
\begin{eqnarray*}
&& a \circ b \in {\cal{A}}\\
%&& a \circ b = b \circ a \\
&& (a \circ b) \circ c= a \circ (b \circ c) \\
&& \left[ a,b \right] \in {\cal{A}} \\
&& \left[ a,b \right]=-[b,a] \\
&& \left[ a, \left[ b,c \right] \right]+\left[ c, \left[ a,b \right] 
\right]+
\left[ b, \left[ c,a \right] \right]=0 \\
&& \left[ a,b \circ c \right] = b \circ \left[ a,c \right]+ \left[ a,b
\right] 
\circ c
\end{eqnarray*} 
\end{definition}

\noindent {\bf{Remark:}}{ \it In the sequel, we will restrict considerations
either  to ``Classical Jordan-Lie'' algebras (hereby referred to as {\rm{CJL}}),
where the  associative product $\circ$ is commutative, or to ``Quantum
Jordan-Lie'' algebras (hereby referred to as {\rm{QJL}}), where the Lie product
$[,]$ is defined through the  associative product  $\circ$ as $[a,b]=(a  \circ b
- b  \circ a)$}

\subsection{Dynamics and algebraic integrability}

>From a purely algebraic point of view we can describe a
Hamiltonian system as a pair $({\cal{A}},H)$ where ${\cal{A}}$ is a
generalized (Classical or Quantum) 
Jordan-Lie algebra and $H$ is a  distinguished element of
${\cal{A}}$ such that the time evolution of any other element $x$ of
$A$ is given by
\be
\dot x=[x,H].
\label{em}
\ee
Our aim is to construct classical and quantum integrable systems.
Within the previous context, this will be achieved by constructing
(Classical or Quantum) 
Jordan-Lie algebras that contain a ``sufficiently large'' abelian
subalgebra  $${\cal{C}}=\{ x_i \in {\cal{A}}\quad / \quad [x_i,x_j]=0
\}$$ and by taking a given element of
${\cal{C}}$ as the Hamiltonian of the system. Under such a Hamiltonian,
(\ref{em}) implies that all the elements of ${\cal{C}}$ are constants
of the motion in involution. In order to ensure complete integrability,
the dimension of the abelian subalgebra should be equal to the number
of degrees of freedom of $H$ under a certain (either symplectic or
quantum-mechanical) realization. In addition, under such physical
realizations, the commuting elements have to be functionally
independent.

\section{A constructive formalism}

We remark that both CJL and QJL are closed under the tensor product,
i.e., if $A$ and $B$ are CJL or QJL, the same holds for $A \otimes B$ 
with the definitions:
\begin{eqnarray*}
&& (a \otimes b) \circ_{A \otimes B} (a' \otimes b')= a \circ_A   a'
\otimes b 
\circ_B b'\\
&& \left[ a \otimes b,a' \otimes b' \right]_{A \otimes B}=[a,a']_A
\otimes b \circ_B b' + 
a' \circ_A  a \otimes [b,b']_B
\end{eqnarray*}
In the sequel, we will skip the subscripts labelling the
spaces we are acting on.

Let us consider a set $\{A_1, \dots, A_N\}$ of CJL or QJL; 
then $A_1 \otimes \cdots \otimes A_N$ is again endowed with a
Jordan-Lie algebra structure. We will take this Jordan-Lie algebra
as our algebra of observables. The following Theorem yields an
algorithmic procedure to obtain CJL or QJL with
large abelian subalgebras (a
Poisson-map version of this construction was already mentioned in
\cite{GMM}). Its proof is obtained by a
straightforward  computation.

\begin{theorem}
Let us assume that two (Classical or Quantum) 
Jordan-Lie algebras $A_1$, $A_2$ are given such that
$A_1$  has a Casimir $C$, and there exists a linear map  
\begin{displaymath}
\phi:A_1 \rightarrow A_1 \otimes A_2
\end{displaymath} 
which is a Jordan-Lie algebra homomorphism:
\bea
&& [\phi(a),\phi(b)]=\phi([a,b]) \quad \forall a,b \in A_1 \cr
&& \phi(a \, b)=\phi(a)\,\phi(b) \quad \forall a,b \in A_1 .
\nonumber
\eea
Let us define the following sequence of homomorphisms
\begin{eqnarray*}
&& \phi^{(2)}\equiv \phi\\
&& \phi^{(i)}=(\phi^{(2)} \otimes \overbrace{id\otimes \dots\otimes
id}^{i-2})
\circ
\phi^{(i-1)}  
\qquad (i=3,\dots,N) \quad
\end{eqnarray*}
where clearly
$$
\phi^{(i)}:A_1 \rightarrow A_1 \otimes 
\overbrace{A_2 \otimes \cdots \otimes A_2}^{i-1}.
$$
Then, the elements $C^{(i)}$ given by
\begin{eqnarray*}
&& C^{(2)} =\phi^{(2)}(C) \otimes \overbrace{1 \otimes \dots \otimes
1}^{N-2}\\ &&\vdots   \\
&& C^{(N-1)}=\phi^{(N-1)}(C) \otimes 1 \\
&& C^{(N)}=\phi^{(N)}(C)
\end{eqnarray*}
commute within $A_1 \otimes 
\overbrace{A_2 \otimes \cdots \otimes A_2}^{N-1}$.
Moreover, 
$$
\conm{\phi^{(N)}(a)}{C^{(i)}}=0\qquad \forall\,a\in
A_1,\qquad (i=2,\dots,N).
$$
\end{theorem}

A relevant example of such a map $\phi$ is given by a coaction mapping
from a Jordan-Lie algebra to another Jordan-Lie Hopf algebra.

\subsection{Integrable systems from comodule algebras}

Let us recall the definition of a coaction of a Hopf algebra $H$ on a
vector space $V$:
\begin{definition}
A (right) coaction of a Hopf algebra $(H,\Delta)$ 
on a vector  space $V$  is a linear map $\phi: V
\rightarrow V\otimes H$ such that 
\begin{displaymath}
(\phi \otimes id) \circ \phi=(id \otimes \Delta) \circ \phi
\end{displaymath}
i.e. if the following diagram is commutative: 
\begin{center}
\setlength{\unitlength}{1cm}
\begin{picture}(6,4)
\put(5,2){$V \otimes H \otimes H$}
\put(.8,2){$V$}
\put(2.5,.8){$V \otimes H$}
\put(2.5,3.3){$V \otimes H$}
\put(3.7,3.2){\vector(4,-3){1.2}}
\put(3.7,1){\vector(4,3){1.2}}
\put(1.2,1.9){\vector(4,-3){1.2}}
\put(1.2,2.3){\vector(4,3){1.2}}
\put(1.35,2.9){$\phi$}
\put(1.35,1.2){$\phi$}
\put(4.45,2.9){$\phi \otimes id$}
\put(4.45,1.2){$id \otimes \Delta$}
\end{picture}
\end{center}
\end{definition}
If $V$ is an algebra, we shall say that $V$ is a {\it $H$-comodule
algebra} if the coaction
$\phi$ is a homomorphism with respect to the product on the algebra
\begin{displaymath}
\phi(ab)=\phi(a)\,\phi(b) \qquad \forall a,b \in V.
\end{displaymath}
Moreover, if $V$ has a Jordan-Lie structure and $H$ is a Lie-Hopf
algebra, $V$ will also be a Jordan-Lie $H$-comodule algebra if:
\begin{displaymath}
\phi([a,b])=[\phi(a),\phi(b)] \qquad \forall a,b \in V
\end{displaymath}

So, given a (Classical or Quantum) Jordan-Lie-Hopf algebra 
with at least one Casimir $C$, if we find a  coaction 
$\phi$ of this Lie-Hopf algebra on a (Classical or Quantum) Jordan-Lie algebra, 
then we have all the  ingredients to apply Theorem $1$.

Note that any Hopf algebra $H$ is a $H$-comodule
algebra with respect to itself, since the coaction map is given by the
coproduct map between $H$ and $H\otimes H$. Therefore, the coalgebra
symmetry approach is just a particular case of the Theorem $1$. Many
examples of integrable systems with this type of coalgebra symmetry have
been already found (see, for instance \cite{BR}-\cite{BRjmp}). 

\section{Examples}

\subsection{Classical systems with a $so(2,1)$ comodule algebra
symmetry}

We consider the Poisson $so(2,2)$ Lie algebra with generators
$\{\Jt,\Jpm,\Nt,\Npm\}$ and Poisson brackets:
\bea
&&\{\Jt,\Jpm\}=\{\Nt,\Npm\}=\pm 2 \Jpm,\nonumber\\
&&\{\Jt,\Npm\}=\{\Nt,\Jpm\}=\pm 2 \Npm,\label{algebraso4}\\
&&\{\Jp ,\Jm \}=\{\Np ,\Nm \}=\Jt,\nonumber\\
&&\{\Jpm,\Nmp\}=\pm \Nt,\quad \{\Ja,\Na\}=0,\ \ m=+,-,3.\nonumber
\eea
The two (second order) Casimir functions for this algebra are:
\bea
&&C_1=\frac 12 \Jt^2 +\frac 12 \Nt^2 + \Jp  \Jm  + \Jm  \Jp  + \Np  \Nm 
+ \Nm 
\Np ,\label{casimirso4i}\\ 
&&C_2=\frac 12 \Jt \Nt + \Jp  \Nm  +\Jm  \Np  .\label{casimirso4ii}
\eea

The algebra $so(2,2)$ can be seen as a $so(2,1)$-comodule algebra
through a coaction
\begin{displaymath}
\phi^{(2)}: so(2,2) \rightarrow so(2,2) \otimes so(2,1)
\end{displaymath}
given by the map
\bea
&\phi^{(2)}(\Jt)=\Jt \otimes 1 + 1 \otimes Y_3 &\qquad
\phi^{(2)}(\Nt)=\Nt
\otimes 1 + 1 \otimes Y_3\cr
 &\phi^{(2)}(\Jp)=\Jp \otimes 1 + 1
\otimes Y_+ &\qquad \phi^{(2)}(\Np)=\Np \otimes 1 + 1 \otimes Y_+\cr
&\phi^{(2)}(\Jm)=\Jm \otimes 1 + 1 \otimes Y_- 
&\qquad
\phi^{(2)}(\Nm)=\Nm \otimes 1 + 1 \otimes Y_-
\nonumber
\eea
where the Lie brackets in $so(2,1)$ are
$$
\{Y_3,Y_{\pm}\}=\pm\,2\, Y_{\pm}\qquad \{Y_+ ,Y_- \}=Y_3
\nonumber
$$
and the  coproduct in $so(2,1)$ is the primitive one
\begin{displaymath}
\Delta(X)=X \otimes 1+ 1 \otimes X \qquad  X=Y_{\pm},Y_3.
\end{displaymath}

It is immediate to define the chain of  homomorphisms 
\begin{displaymath}
\phi^{(i)}=(\phi^{(2)} \otimes \overbrace{id\otimes \dots\otimes
id}^{i-2})
\circ
\phi^{(i-1)}  
\qquad (i=3,\dots,N) .
\end{displaymath}
As a consequence, the set of $2\,N$ operators
\begin{eqnarray}
&&C_1^{(i)}=\phi^{(i)}(C_1)\otimes \overbrace{1 \otimes 1 \otimes
\cdots \otimes 1}^{N-i}
\label{cas1}\\
&&C_2^{(i)}=\phi^{(i)}(C_2)\otimes \overbrace{1 \otimes 1 \otimes
\cdots \otimes 1}^{N-i}
\label{cas2} 
\end{eqnarray}
are in involution with all the elements of the
subalgebra $\phi^{(N)}(so(2,2))$ within 
$so(2,2) \otimes so(2,1)^{\otimes (N-1)}$.

In order to get Classical Mechanical systems we can consider
the symplectic realization $D$ of $so(2,2)$ given by
\bea
&&D(\Jt)= p_1 q_1 + p_2 q_2\cr
&&D(\Jp)=\frac{1}{2}(p_1^2 +p_2^2) + \frac{a_{12}}{(q_1 + q_2)^2} +
\frac{b_{12}}{(q_1 - q_2)^2}  \cr 
&&D(\Jm)=-\frac{1}{2}(q_1^2 +q_2^2)\cr 
&&D(\Nt)=(p_1 q_2 + p_2 q_1) \label{spsoa}\\
&&D(\Np)=p_1 p_2 + \frac{a_{12}}{(q_1 + q_2)^2} -
\frac{b_{12}}{(q_1 - q_2)^2}\cr
&&D(\Nm)=- q_1 q_2
\nonumber
\eea
It is readily checked that the functions (\ref{spsoa}) close the
$so(2,2)$ Poisson algebra. This symplectic realization is
characterized by the Casimirs, whose values are
$D(C_1)=-(a_{12}+b_{12})$ and
$D(C_2)=-(a_{12}-b_{12})/2$. (Note that $D(\Jp)$ is just the
two-body rational Calogero-Moser  Hamiltonian).

Secondly, we consider the symplectic realization $S$ of $so(2,1)$
defined by :
$$
S(Y_3)=p_3 q_3\qquad
S(Y_+)=\frac{p_3^2}{2}+ \frac{c_{3}}{q_3^2}\qquad
S(Y_+)=-\frac{q_3^2}{2}.
$$

By using these realizations, the image of the generators under the
coaction is
\bea
&&\phi^{(2)}(\Jt)= p_1 q_1 + p_2 q_2 + p_3 q_3\cr
&&\phi^{(2)}(\Jp)=\frac{1}{2}(p_1^2 +p_2^2) + \frac{a_{12}}{(q_1 +
q_2)^2} + \frac{b_{12}}{(q_1 - q_2)^2} +\frac{p_3^2}{2}+
\frac{c_{3}}{q_3^2}  \cr 
&&\phi^{(2)}(\Jm)=-\frac{1}{2}(q_1^2
+q_2^2) -\frac{q_3^2}{2}\cr  &&\phi^{(2)}(\Nt)=p_1 q_2 + p_2 q_1 +p_3
q_3\label{spso}\\ &&\phi^{(2)}(\Np)=p_1 p_2 + \frac{a_{12}}{(q_1 +
q_2)^2} -
\frac{b_{12}}{(q_1 - q_2)^2} + \frac{p_3^2}{2}+ \frac{c_{3}}{q_3^2}\cr
&&\phi^{(2)}(\Nm)=- q_1 q_2 -\frac{q_3^2}{2}
\nonumber
\eea
and the image of the Casimir functions
$C_1$ and
$C_2$ is
nomore a constant:
\bea
&& C_1^{(2)}:=(D \otimes S)(C_1)= -(a_{12}+b_{12}+2c_3)+p_3 q_3 (p_1
+p_2)(q_1+q_2)-\cr
&& -(\frac{1}{2} p_3^2+\frac{c_3}{q_3^2})(q_1+q_2)^2- 
q_3^2 \left[ \frac{1}{2} (p_1+
p_2)^2+\frac{2 a_{12}}{(q_1+q_2)^2} \right] 
\cr  
&& C_2^{(2)}:=(D \otimes S)(C_2)=
\frac{1}{2}\,C_1^{(2)} + b_{12}
\eea
meaning that both Casimirs give rise to functionally dependent images.
Therefore, the function $C_1^{(2)}$  and the image of two
Poisson-commuting generators of $so(2,2)$  (for instance, we could take
$\phi(\Jp)$ and $\phi(N_+)$) give us a complete family of Poisson
commuting observables and allow us to construct a three-body completely
integrable Hamiltonian system.

The $N$-th image would be given by
\bea
&&\phi^{(N)}(\Jt)= p_1 q_1 + p_2 q_2 + \sum_{k=3}^{N+1}{p_k q_k}\cr
&&\phi^{(N)}(\Jp)=\frac{1}{2}(p_1^2 +p_2^2) + \frac{a_{12}}{(q_1 +
q_2)^2} + \frac{b_{12}}{(q_1 - q_2)^2} +
\sum_{k=3}^{N+1}{\big(\frac{p_k^2}{2}+
\frac{c_{k}}{q_k^2}\big)}  \cr 
&&\phi^{(N)}(\Jm)=-\frac{1}{2}(q_1^2
+q_2^2) -\sum_{k=3}^{N+1}{\frac{q_k^2}{2}}\cr
&&\phi^{(N)}(\Nt)=p_1 q_2 +
p_2 q_1 + \sum_{k=3}^{N+1}{p_k q_k}\label{spso}\\
&&\phi^{(N)}(\Np)=p_1 p_2 +
\frac{a_{12}}{(q_1 + q_2)^2} -
\frac{b_{12}}{(q_1 - q_2)^2} + \sum_{k=3}^{N+1}{\big(\frac{p_k^2}{2}+
\frac{c_{k}}{q_k^2}\big)}\cr
&&\phi^{(N)}(\Nm)=- q_1 q_2 -\sum_{k=3}^{N+1}{\frac{q_k^2}{2}}
\nonumber
\eea
where we have used the notation
$$
p_i=\overbrace{1 \otimes \cdots \otimes 1}^{i-1} \otimes p \otimes 
\overbrace{1 \otimes \cdots \otimes 1}^{N-i}\qquad
q_i=\overbrace{1 \otimes \cdots \otimes 1}^{i-1} \otimes q \otimes 
\overbrace{1 \otimes \cdots \otimes 1}^{N-i}.
$$  

A complete family of Poisson commuting observables in the $(N+1)-$body
case is provided by the symplectic realizations of the
images of the Casimirs
\bea
&& C_1^{(M)}=-(a_{12}+b_{12}+2 \sum_{k=3}^{M+1} c_k)+ \sum_{k=3}^{M+1} 
p_k q_k (p_1 +p_2)(q_1+q_2)- \cr
&& -\sum_{j=3}^{M+1} \sum_{k > j}^{M+1} (q_j p_k-p_k q_j)^2-
 \sum_{k=3}^{M+1}
(\frac{1}{2} p_k^2+\frac{c_k}{q_k^2})(q_1+q_2)^2- \cr
&& 2 \sum_{j,k=3}^{M+1}  q_j^2 \frac{c_k}{q_k^2} - \sum_{k=3}^{M+1} q_k^2 
\left[ \frac{1}{2}
(p_1+ p_2)^2+\frac{2 a_{12}}{(q_1+q_2)^2} \right]  \qquad M=2,\dots,N
\nonumber
\eea
together with two commuting elements (for instance,
$\phi^{(N)}(\Jp)$ and $\phi^{(N)}(\Np)$).

\subsection{Poisson-Schr\"odinger comodule algebras}

Certain subalgebras of Hopf algebras provide
examples of comodule algebras which are useful in order
to construct new integrable systems. In particular, let
$(B,\Delta_B)$ be a Hopf algebra and let $A\subset B$
be a subalgebra of $B$ (as an algebra, but not as a Hopf algebra: this
means that the coproduct $\Delta_B$ of the elements of $A$ contain
some elements which do not belong to $A\otimes A$). In this case it is
straightforward to prove that, if there exists a basis in
$B$ such that the coproduct $\Delta_B$ of the elements of $A$ is of the
form
\be
\Delta_B (X)=\sum_{\alpha}{X_\alpha \otimes Y_\alpha} \qquad \forall
X\in A, \qquad 
\mbox{where}\quad X_\alpha\in A\quad \mbox{and}\quad Y_\alpha\in B
\label{cond}
\ee
then $A$ is a $B$-comodule algebra through the coaction
$\phi:A\rightarrow A\otimes B$ given by
\be
\phi(X):=\Delta_B (X) \qquad 
\forall X\in A.
\ee
In the non-commutative case, this situation is quite
usual within the quantum algebra approach to discrete symmetries 
(see, for instance, \cite{BHNN} and \cite{Herr}). In the following, we
will consider as $B$ the Poisson analogues of two non-standard
deformations of the Schr\"odinger algebra introduced in \cite{BHNN}. The
associated (``deformed") coactions on a $gl(2)$ subalgebra (that plays
the role of
$A$) will give rise to two new integrable deformations of the isotropic
oscillator in
$N$-dimensions.

\subsubsection{The ``space-type" comodule Schr\"odinger system}

Let us consider the following Schr\"odinger Poisson-Hopf algebra ${\cal
S_\sigma}\equiv B$:
\be
\begin{array}{llll}
 \{\luisD,\luisP\}=-\luisP \quad   &\{\luisD,\luisK\}=\luisK \quad
&\{\luisK,\luisP\}= \luisM \quad &\{\luisM,\,\cdot\,\}=0\cr
 \{\luisD,\luisH\}=-2\luisH \quad  &\{\luisD,\luisC\}=2\luisC \quad
&\{\luisH,\luisC\}=\luisD \quad &\{\luisH,\luisP\}=0 \cr
 \{\luisP,\luisC\}= - \luisK  \quad  &
 \{\luisK,\luisH\}= \luisP\quad  &\{\luisK,\luisC\}=0  
  \quad  &
\end{array}
\label{hd}
\ee
and the deformed coproduct compatible with these Poisson brackets is
found to be 
\bea
&&\Delta(\luisM)=1\otimes \luisM + \luisM\otimes 1\cr
&&\Delta(\luisH)=1\otimes \luisH +
\luisH\otimes (1 + \sigma \luisP)^2\cr
&&\Delta(\luisD)=1\otimes \luisD + \luisD\otimes  \frac{1}{1 +
\sigma \luisP} - 
\frac{1}2 \luisM\otimes \frac{\sigma \luisP}{1 +
\sigma \luisP}\cr
&&\Delta(\luisC)=1\otimes \luisC + \luisC\otimes
\frac{1}{(1 + \sigma \luisP)^2} + \sigma   \luisD'\otimes
\frac{1}{1 + \sigma \luisP}\,\luisK + \frac{\sigma^2}{2}(\luisD')^2
\otimes
\frac{\luisM}{(1 +
\sigma \luisP)^2}\qquad\label{ss}\\
&&\Delta(\luisP)=1\otimes \luisP + \luisP\otimes 1 + \sigma
\luisP\otimes \luisP\cr
&&\Delta(\luisK)=1\otimes
\luisK + \luisK\otimes  \frac{1}{1 + \sigma \luisP} +\sigma
\luisD'\otimes \frac{\luisM}{1 + \sigma \luisP}
\nonumber
\eea
where $\luisD'=\luisD+\frac 12 \luisM$.  
This algebra is the Poisson analogue of the ``discrete space"
non-standard quantum Schr\"odinger algebra of \cite{BHNN}, where
$\sigma$
is the deformation parameter (if $\sigma\rightarrow 0$, we recover the
primitive coproduct 
$\Delta(X)=X \otimes 1+ 1 \otimes X$).

If we consider the Poisson-$gl(2)$ subalgebra generated by
$\{\luisM,\luisH,\luisD,\luisC\}$ as our subalgebra $A$, it is
straightforward to check that the Poisson-$gl(2)$ subalgebra fulfills
the condition (\ref{cond}). As a consequence, $gl(2)$ is a
(Poisson)-${\cal
S_\sigma}$ comodule algebra and (\ref{ss}) can be used to define a
(right) coaction map 
$
\phi^{(2)}:gl(2)\rightarrow gl(2)\otimes {\cal
S_\sigma}
$ 
in the form:
\be
\phi^{(2)}(X):=\Delta (X) \qquad 
 X\in \{\luisM,\luisH,\luisD,\luisC\}.
\label{cos}
\ee

A symplectic realization of ${\cal S_\sigma}$ in terms of a pair of
canonical coordinates is given by:
\bea
&& 
S(\luisC)=\frac{q_1^2}{2}\qquad
S(\luisH)=\frac{p_1^2}{2}\qquad
S(\luisD)=-p_1\,q_1\cr
&&
S(\luisM)=\lambda_1^2\qquad\!\!\!
S(\luisK)=\lambda_1\,q_1\quad\,
S(\luisP)=\lambda_1\,p_1
\label{sr}
\eea
Under this one-particle symplectic realization, the $gl(2)$ Casimir
function
\be
C_{A}=\frac{1}{4}\,\luisD^2 - \luisH\,\luisC
\label{ca}
\ee
vanishes $S(C_{A})=0$ (note that $\luisM$ is also a trivial central
element).

If we take as the Hamiltonian on $gl(2)$ the function
\be
H=\luisH + \luisC
\ee
the symplectic realization $S$ will give us the one dimensional
harmonic oscillator:
$$
H^{(1)}:=S(H)=S(\luisH)+S(\luisC)=\frac{p_1^2}{2}+\frac{q_1^2}{2}.
$$
Now, by using the coaction map (\ref{cos}) we can define a Hamiltonian
function on $gl(2)\otimes {\cal S_\sigma}$:
$$
\phi^{(2)}(H)=\phi^{(2)}(\luisH)+\phi^{(2)}(\luisC)=\Delta(\luisH)
+\Delta(\luisC).
$$
This Hamiltonian can be expressed in terms of canonical
coordinates by taking the symplectic realization
$S\otimes S$ of (\ref{ss}). It reads:
\bea
&& \back\back H^{(2)}_\sigma=(S\otimes S)(\phi^{(2)}(H))=(S\otimes
S)(\Delta(\luisH) +\Delta(\luisC))\cr
&& \qquad\, \back\back =\frac{1}{2}(p_1^2 + p_2^2)+\frac{q_2^2}{2}
+ \frac{q_1^2}{2(1+\sigma\,\lambda_2\,p_2)^2}
\cr
&& \qquad\quad + \sigma\,\lambda_2\left( p_1^2\,p_2 +
\frac{q_2(\lambda_1^2 - 2
q_1\,p_1)}{2(1+\sigma\,\lambda_2\,p_2)}\right)
+
\sigma^2\,\lambda_2^2\,\left(\frac12 p_1^2\,p_2^2 + 
\frac{(\lambda_1^2 - 2
q_1\,p_1)^2}{8(1+\sigma\,\lambda_2\,p_2)^2}   \right).
\label{h2s}
\eea
This Hamiltonian is just an integrable deformation of the
two-dimensional
isotropic oscillator, since $\lim_{\sigma\to
0}{H^{(2)}}=\tfrac{1}{2}(p_1^2 + p_2^2)+\tfrac12 (q_1^2 +
q_2^2)$. The integral of  motion is obtained as
the coaction of the $gl(2)$ Casimir (\ref{ca}):
\bea
&& \back\back C^{(2)}_\sigma=(S\otimes S)(\phi^{(2)}(C_{A}))=
(S\otimes S)(\frac{1}{4}\,\Delta(\luisD)^2 - \Delta(\luisH))
\label{c2s} \\
 && \qquad\, \back\back 
=-\frac{\{ 2(p_2 q_1-p_1 q_2) +
\sigma p_1 (2 p_1 q_1- 4 p_2 q_2-\lambda_1^2) -
\sigma^2 \lambda_2^2 p_1 p_2 (-2 p_1 q_1+2  p_2 q_2+\lambda_1^2)
\}^2}{16\,(1+\sigma \lambda_2 p_2)^2}
\nonumber
\eea
As expected, the limit $\sigma\rightarrow 0$ of (\ref{c2s}) is just
$-(p_2 q_1-p_1 q_2)^2/4$.

By following Theorem 1, further iterations of the coaction map would 
provide the corresponding integrable deformation of the isotropic
oscillator in an arbitrary dimension.

\subsubsection{The ``time-type" comodule Schr\"odinger system}

Another integrable deformation of the isotropic oscilator can be
obtained
through the Poisson version  of the ``discrete space" quantum
Schr\"odinger algebra ${\cal S_\tau}$ \cite{BHNN}. In this case, the
deformed coproduct compatible with the Poisson-Schr\'odinger algebra
(\ref{hd}) is:
\bea
&&\Delta(\luisM)=1\otimes \luisM +
\luisM\otimes 1\cr
&&\Delta(\luisH)=1\otimes \luisH + \luisH\otimes 1 +
\tau \luisH\otimes \luisH\cr
&&\Delta(\luisD)=1\otimes \luisD
 + \luisD\otimes  \frac{1}{1 + \tau \luisH}
 - \frac{1}2 \luisM\otimes \frac{\tau \luisH}{1 + \tau
\luisH}\cr 
&&\Delta(\luisC)= 1\otimes \luisC + \luisC\otimes 
\frac{1}{1 + \tau
\luisH}   - \frac {\tau}{2} \luisD'\otimes 
  \frac{1}{1 + \tau \luisH}
\luisD + \frac{\tau^2}4 (\luisD')^2\otimes
\frac{\luisH}{(1 + \tau \luisH)^2}\cr
&&\Delta(\luisP)=1\otimes \luisP +
\luisP\otimes (1 + \tau \luisH)^{1/2}\cr
&&\Delta(\luisK)=1\otimes
\luisK + \luisK\otimes 
\frac{1}{(1 + \tau \luisH)^{1/2}} +\frac{\tau}{2} 
\luisD'\otimes \frac{\luisP}{1 + \tau \luisH}.
\label{kd}
\eea

Once again, we can consider $gl(2)$ as the subalgebra $A$, since the
coproduct (\ref{kd}) of the generators
$\{\luisM,\luisH,\luisD,\luisC\}$ defines a coaction $
\phi^{(2)}:gl(2)\rightarrow gl(2)\otimes {\cal
S_\tau}
$.
By taking as the Hamiltonian on $gl(2)$ the same function
$
H=\luisH + \luisC
$
and by considering the symplectic realization (\ref{sr})
we obtain:
\bea
&& \back\back H^{(2)}_\tau=(S\otimes S)(\phi^{(2)}(H))=
=(S\otimes
S)(\Delta(\luisH) +\Delta(\luisC))\cr
&& \qquad\, \back\back
=\frac{1}{2}(p_1^2 + p_2^2)+\frac{q_2^2}{2}
+ \frac{q_1^2}{2+\tau\,p_2^2}
\cr
&& \qquad\quad + \tau\,\left(\frac14 p_1^2\,p_2^2 +
\frac{p_2\,q_2(\lambda_1^2 - 2
q_1\,p_1)}{2(2+\tau\,p_2^2)}\right)
+
\tau^2\,\left( 
\frac{p_2\,(\lambda_1^2 - 2
\,q_1\,p_1)}{8(2+\tau\,p_2^2)}   \right)^2.
\label{h2s}
\eea
In this case the integral of the motion reads:
\bea
&& \back\back C^{(2)}_\tau=(S\otimes S)(\phi^{(2)}(C_{A}))=
(S\otimes S)(\frac{1}{4}\,\Delta(\luisD)^2 - \Delta(\luisH))
\label{c2s} \\
 && \qquad\, \back\back 
=-\frac{\left\{4(p_1 q_2-p_2 q_1) -
2\,\tau\,p_1^2\,p_2\,q_1 +
\tau\,p_1\,p_2\,(2\,p_2\,q_2+\lambda_1^2)
\right\}^2}{32\,(2+\tau \, p_2^2)}
\nonumber
\eea
As in the previous case, the procedure can be extended to arbitrary
dimensions. However a compact and explicit expression for
the $N$-th coaction $\phi^{(N)}$ is still lacking.

%%%%%%%%%%%%%%%%%%%%%%%%%%%%%%%%%%%%%%%%%%%%%%%%%%%%%%%%%%%%%%%%%%%%%%

\subsection{$q$-oscillator systems}

Let us now consider as our vector space $V$ the ``$q$-oscillator 
algebra'' $A_q$ defined by the commutation relations
\cite{Bi}-\cite{Kul}:
\begin{eqnarray*}
\big[ N,A \big]&=&-A\\
\left[ N,A^+ \right]&=&A^+\\
\left[ A,A^+ \right]&=&q^{-2N}
\end{eqnarray*}
and as our coalgebra $H$ we shall take the quantum algebra $su_q(2)$
\cite{KR} defined by the commutation rules: 
\begin{eqnarray*}
\left[ J,X_{\pm} \right]&=&X_{\pm}\\
\left[ X_{+},X_{-} \right]&=&\frac{q^{2J}-q^{-2J}}{q-q^{-1}}=[2\,J]_q
\end{eqnarray*}
and Casimir operator
$$
L_q=[J]_q\,[J-1]_q + X_+\,X_-,
\nonumber
$$
where $[j]_q\,[j+1]_q$ is the eigenvalue of $L_q$ in the $(2\,j+1)$
dimensional irreducible representation of $su_q(2)$. The coaction
$\phi^{(2)}$ is defined by \cite{Kul}:
\begin{eqnarray*}
\phi^{(2)}(N)&=&N \otimes 1 + 1 \otimes J\\
\phi^{(2)}(A)&=& A \otimes q^{J}+ (q-q^{-1})^{\frac{1}{2}} q^{-N}
\otimes X_{-}\\
\phi^{(2)}(A^+)&=& A^+ \otimes q^{J}+ (q-q^{-1})^{\frac{1}{2}} q^{-N}
\otimes X_{+}
\end{eqnarray*}

It is easy to check that this coaction turns $A_q$ into a (Jordan-Lie)
$su_q(2)$ comodule-algebra. On the other hand,
$A_q$ is equipped with the Casimir 
\begin{displaymath}
C_q=A^+ A - \frac{q^{-2N}-1}{q^{-2}-1}
\end{displaymath}
which is not invariant under the coaction $\phi^{(2)}$.

\subsubsection{Quantum Hamiltonian}

If we take as Hamiltonian \cite{Kul,ChaKul}
$$
H=A^+\,A
$$
its image under the first coaction will be
$H^{(2)}=\phi^{(2)}(H)=\phi^{(2)}(A^+)\,\phi^{(2)}(A)$ and reads
$$
H^{(2)}=
A^+\,A\,q^{2J}+(q-q^{-1})\,q^{-2N}\,X_{+}\,X_{-}
+(q-q^{-1})^{\frac{1}{2}}
\,q^{-N+J}\,\{q^{-1}\,A\,X_{+} + q\,A^+\,X_{-} \}.
$$
The coaction of the Casimir $\phi^{(2)}(C_q)$ yields the constant
of the motion $C^{(2)}$:
$$
\phi^{(2)}(C_q)=\phi^{(2)}\left(A^+ A -
\frac{q^{-2N}-1}{q^{-2}-1}\right)=\phi^{(2)}(H) -
\phi^{(2)}\left(\frac{q^{-2N}-1}{q^{-2}-1}\right),
$$
namely
$$
C^{(2)}=H^{(2)}-\frac{q^{-2(N+J)}-1}{q^{-2}-1}.
$$
Since $C^{(2)}$ is a constant of the motion, the
conservation of $(N+J)$ -the total number of excitations- follows.

In fact, the coaction $\phi^{(2)}$ can be thought of a homomorphism between
$A_q$ and $A_q\otimes A_q\otimes A_q$, since $su_q(2)$ can be rewritten
in terms of two $q$-oscillators through the $q$-analogue of the
Jordan-Schwinger transformation \cite{Jo,Sch}:
\bea
&& J=\frac12 (N_1-N_2) \cr
&& X_+=B^+\,q^{N_1/2}\,q^{N_2/2}\,C \label{qJS}\\
&& X_{-}=q^{N_1/2}\,B\,C^+\,q^{N_2/2} 
\nonumber
\eea
where $\{N_1,B,B^+\}$ and $\{N_2,C,C^+\}$ are two copies of the
the $C_q=0$ irreducible
representation of the algebra $A_q$ (note that the
representation theory of the
$q$-oscillator is much more complex than the oscillator one; this should
be taken into account because different models could be obtained for
each class of representations -see, for instance, \cite{qosc}). In terms
of this Jordan-Schwinger realization, the $H^{(2)}$ Hamiltonian is
rewritten as:
\bea
&& \back\back H^{(2)}=
A^+\,A\,q^{N_1-N_2}
+(q-q^{-1})\,q^{-2N}\,\left([j]_q\,[j+1]_q - [\frac12
(N_1-N_2)]_q\,[\frac12 (N_1-N_2 -2)]_q\right)\cr && \qquad\qquad
+(q-q^{-1})^{\frac{1}{2}}
\,q^{-\frac12
(2\,N-3\,N_1+N_2+1)}\,\{q^{-1}\,A\,B^+\,C +
q\,A^+\,B\,C^+ \}.
\nonumber
\eea
We thus have a $q$-deformed three-wave interaction
Hamiltonian that preserves the number operator $N+N_1 -N_2$ and whose
limit $q\rightarrow 1$ is just $A^+\,A$.

The $k$-dimensional model is obtained through the $k$-th iteration of
the coaction map. In particular, the $k$-dimensional Hamiltonian reads:
\bea
&& \back\back H^{(k)}=
\phi^{(k)}(A^+\,A)=(A^+\,A)\,
\Delta^{(k-1)}(q^{2J})+
(q-q^{-1})\,q^{-2N}\,\Delta^{(k-1)}(X_{+}\,X_{-})
\cr
&&\qquad  +(q-q^{-1})^{\frac{1}{2}} 
\left\{(q^{-N}\,A)\,\Delta^{(k-1)}(X_{+}\,q^{J}) +
(A^+\,q^{-N})\,\Delta^{(k-1)}(q^{J}X_{-}) \right\},
\nonumber
\eea
where $\Delta^{(k-1)}$ is the $(k-1)$-th coproduct in $su_q(2)$.
This Hamiltonian is completely integrable, since it commutes with the
$m$-th order Casimirs, which are obtained as the corresponding images
under the $m$-th coaction:
$$
C^{(m)}=\phi^{(m)}(C_q)=H^{(m)}-\phi^{(m)}\left(
\frac{q^{-2(N)}-1}{q^{-2}-1}\right)=H^{(m)}-\left(
\frac{q^{-2\,\phi^{(m)}(N)}-1}{q^{-2}-1}\right).
$$
By construction, all these integrals are in involution
$$
\conm{C^{(l)}}{C^{(p)}}=0, \qquad 2\leq l < p \leq k.
$$
In particular, the commutation rule $\conm{H^{(k)}}{C^{(k)}}=0$ implies
that the $\phi^{(k)}(N)$ operator commutes with the Hamiltonian.
Finally, note that through the Jordan-Wigner transformation (\ref{qJS})
the
$k$-th order Hamiltonian can be seen as a system of $(2\,k +1)$
interacting
$q$-oscillators.

\subsubsection{Classical model}

The previous model can be translated in classical mechanical terms by
an appropriate definition of the ``$q$-oscillator Poisson algebra'' that
(to our knowledge) is a new one. The classical Poisson algebra
$A_q$ is defined by
\bea
\pois{N}{A}&=&-A\cr
\pois{N}{A^+}&=&A^+ \label{po}\\
\pois{A}{A^+}&=&e^{-2 z N}
\nonumber
\eea
and as our coalgebra $H$ we will take the $su_q(2)$ Poisson coalgebra
given by
\begin{eqnarray*}
\pois{J}{X_{\pm}}&=&X_{\pm}\\
\pois{X_{+}}{X_{-}}&=&\frac{e^{2 z J}-e^{-2 z J}}{2\,z}
\end{eqnarray*}
The coaction $\phi^{(2)}$ compatible with (\ref{po}) is
\begin{eqnarray*}
\phi^{(2)}(N)&=&N \otimes 1 + 1 \otimes J\\
\phi^{(2)}(A)&=& A \otimes e^{z J}+ \sqrt{2\,z} \,e^{-z N} \otimes
X_{-}\\
\phi^{(2)}(A^+)&=& A^+ \otimes e^{z J}+ \sqrt{2\,z} \,e^{-z N} \otimes
X_{+}.
\end{eqnarray*}
Since $A_q$
is equipped with the Casimir function,
\begin{displaymath}
C_q=A^+ A - \frac{1-e^{-2 z N}}{2\,z}
\end{displaymath}
a $k$ dimensional abelian subalgebra inside
$\phi^{(k)}(A_q)$ can be constructed through the approach described in
the previous section. The results are similar to the ones for the
``quantum"
$q$-oscillator algebra, and the Jordan-Schwinger transformation is
also valid in the Poisson case (with $q=e^z$). Let us also mention that
a symplectic realization of the Poisson algebra (\ref{po})
corresponding to the value $C_q=0$ is given by the relations:
\bea
&& N= p_1\,q_1 \cr
&& A^+= p_1 \cr
&& A= \frac12
\left(\frac{1-e^{-2\,z\,p_1\,q_1}}{z\,p_1\,q_1}\right)\,q_1.
\nonumber
\eea
Under this realization, $H^{(k)}$ is an integrable hamiltonian with $k$
degrees of freedom and integrals of the motion $C^{(l)}$
($l=2,\dots,k$).

\subsection{Integrable Hamiltonians and the Reflection Equation algebra}

The RE algebra $\cal{A}$ \cite{KS} provides a representative example of
a four-dimensional comodule algebra with two Casimir elements. The
commutation rules of the RE algebra are:
\bea
&\conm{\alpha}{\beta}=(q-q^{-1})\,\alpha\,\gamma    \qquad
&\qquad {\alpha}\,{\gamma}=q^{2}\,\gamma\,\alpha \cr
&\conm{\alpha}{\delta}=(q-q^{-1})\,(q\,\beta+\gamma)\,\gamma \qquad   
&\qquad \conm{\beta}{\gamma}=0 \cr
&\conm{\beta}{\delta}=(q-q^{-1})\,\gamma\,\delta  \qquad     
&\qquad {\gamma}\,{\delta}=q^{2}\,\delta\,\gamma
\nonumber
\eea
and the two central elements of $\cal{A}$ are:
$$
c_1=\beta -q\,\gamma \qquad\qquad c_2=\alpha\,\delta-q^{2}\beta\,\gamma.
$$

The RE algebra is a $GL_q(2)$-comodule algebra. We recall the $GL_q(2)$
commutation rules of a $GL_q(2)$ element $T$
$$
T=\pmatrix{a & b \cr
c & d} 
$$
which read
\bea
&&a\,b=q\,b\,a  \qquad
\qquad a\,c=q\,c\,a  \qquad
\qquad \conm{a}{d}=(q-q^{-1})\,b\,c \cr
&&b\,d=q\,d\,b  \qquad
\qquad c\,d=q\,d\,c  \qquad
\qquad \conm{b}{c}=0. 
\nonumber
\eea
The Hopf algebra structure in $GL_q(2)$ is given by the matrix
multiplication of two group elements:
$$
\Delta(T)=\pmatrix{\Delta (a) & \Delta (b) \cr
\Delta (c) & \Delta (d)}=T_1\cdot T_2=\pmatrix{a_1 & b_1 \cr
c_1 & d_1}\cdot \pmatrix{a_2 & b_2 \cr
c_2 & d_2}
$$

If we rewrite the generators of the RE algebra in
matrix form
$$
K=\pmatrix{\alpha & \beta \cr
\gamma & \delta} 
$$
the (left) coaction map $\phi^{(2)}:{\cal A}\rightarrow
GL_q(2)\otimes \cal{A}$ will be given by
$$
\phi^{(2)}(K)=\pmatrix{\phi^{(2)}(\alpha) & \phi^{(2)}(\beta) \cr
\phi^{(2)}(\gamma) & \phi^{(2)}(\delta)}=T\cdot K \cdot T^t
$$
where $T^t$ is the transpose of $T$. Therefore, the $k$-th coaction map
$$
\phi^{(k)}:{\cal A}\rightarrow \overbrace{GL_q(2)\otimes GL_q(2)\otimes
\dots GL_q(2)}^{k-1}\otimes \cal{A}
$$
will be given by
$$
\phi^{(k)}(K)=\pmatrix{\phi^{(k)}(\alpha) & \phi^{(k)}(\beta) \cr
\phi^{(k)}(\gamma) & \phi^{(k)}(\delta)}=
\left\{\prod_{l=1}^{k-1}{T_{k-l}}
\right\}
\cdot K \cdot
\left\{\prod_{l=1}^{k-1}{{T_l}^t}\right\}
$$
where
$$
T_l=\pmatrix{a_l & b_l \cr
c_l & d_l} \qquad\qquad 
{T_l}^{t}=\pmatrix{a_l & c_l \cr
b_l & d_l}.
$$

If we consider as Hamiltonian any function on $\cal{A}$
$$
H=H(\alpha,\beta,\gamma,\delta)
$$
the integrability of the Hamiltonian defined as
$$
H^{(m)}=\phi^{(m)}(H(\alpha,\beta,\gamma,\delta))=
H(\phi^{(m)}(\alpha),\phi^{(m)}(\beta),\phi^{(m)}(\gamma),
\phi^{(m)}(\delta))
$$
is guaranteed by the $2(m-1)$ images of the casimir functions $c_1$ and
$c_2$, that can be written in the following factorized form:
\bea
&& c_1^{(k)}=\phi^{(k)}(c_1)=\left\{\prod_{l=1}^{k-1}{\mbox{det}_q\,
T_l}
\right\}\,c_1 \qquad\quad 2\leq k \leq m\cr
&&
c_2^{(k)}=\phi^{(k)}(c_2)=\left\{\prod_{l=1}^{k-1}{(\mbox{det}_q\,
T_l)^2}
\right\}\,c_2 \qquad 2\leq k \leq m
\nonumber
\eea
where the $q$-determinant on each $GL_q(2)$ copy is given by
$$
\mbox{det}_q\,T_l=a_l\, d_l - q\,b_l\,c_l.
$$
It can be easily proven that
\bea
&& \conm{H^{(m)}}{c_1^{(k)}}=0 \qquad \conm{H^{(m)}}{c_2^{(p)}}=0 \qquad
 2\leq k,p
\leq m \cr
&& \conm{c_1^{(l)}}{c_1^{(n)}}=0 \qquad \conm{c_2^{(l)}}{c_2^{(n)}}=0
 \qquad 2\leq l<n
\leq m\cr
&&
\conm{c_1^{(k)}}{c_2^{(p)}}=0 \qquad k,p=2,\dots, m.
\nonumber
\eea

In general, coaction-induced integrable systems on
quantum homogeneous spaces can be constructed in the same
way.

\section{Concluding remarks}

A further generalization of the construction given in Theorem 1 is the
following. Let us consider a set
$\{A_1,
\dots, A_N\}$ of (Classical or Quantum)
Jordan-Lie algebras, then $A_1 \otimes \cdots \otimes
A_N$ is again endowed with a Jordan-Lie algebra structure. Let us
suppose that we are able to find a set of Jordan-Lie subalgebras
$\{B_1, \dots, B_N\}$ such that 
\begin{displaymath}
B_1 \subset A_1 \otimes A_2 \qquad B_{i+1} \subset B_i \otimes A_{i+2}
\quad
i=1, \dots, N-2.
\end{displaymath}
Let us also assume that a Casimir element $C_{B_i} \in B_i$  with
respect with the bracket on $B_i$ exists for any $i$. If we denote 
\begin{eqnarray*}
\begin{array}{l}
C_1=C_{B_1} \otimes \overbrace{1 \otimes \dots \otimes 1}^{N-2}\\
\vdots   \\
C_{N-2}=C_{B_{N-2}} \otimes 1 \\
C_{N-1}=C_{B_{N-1}}
\end{array}
\end{eqnarray*}
the following Theorem holds:
\begin{theorem}
The set $\{C_1,\dots,C_{N-1}\}$ are mutually commuting elements in $A_1
\otimes
\cdots \otimes A_N$ and central elements for $B_{N-1}$.
\end{theorem}
{\bf{Proof:}}
\vspace{.2cm}

\noindent We have the chain of inclusions:
\begin{displaymath}
B_{i} \subset B_{i-1} \otimes A_{i+1} \subset B_{i-2} \otimes A_i 
\otimes A_{i+1}
\subset \cdots \cdot \cdot\subset B_1 \otimes A_3 \otimes \cdots 
\otimes A_{i+1}  
\end{displaymath}
so that $(i\leq j)$
\begin{eqnarray*}
&& \back\back [C_i,C_j] \in  
 [C_i \otimes \overbrace{1 \otimes \cdots \otimes 1}^{j-i}
\otimes \overbrace{1 \otimes \cdots \otimes 1}^{N-j}, B_i \otimes 
A_{i+1}
\otimes \cdots \otimes A_{j+1} \otimes \overbrace{1 \otimes \cdots 
\otimes 1}^{N-j}]= 0\\
&& [C_i, B_{N-1}] \in [C_i \otimes \overbrace{1 \otimes \cdots \otimes
1}^{N-i}, B_i \otimes  A_{i+1} \otimes \cdots \otimes A_{N}]=0
\hskip3,5cm \Box
\end{eqnarray*}

\noindent Hence,  for  an arbitrary element $H \in B_{N-1}$, the set $\{
H, C_2,
\dots, C_{N-1} \}$ yields an abelian subalgebra of $A_1 \otimes \cdots
\otimes A_N$ of dimension $N$.

On the other hand, the freedom to choose different homomorphisms in
each step of the ``duplication" process should be also mentioned as
another possible generalization of the comodule algebra approach. In
other
words, it would be possible to define the chain of homomorphisms without
making use of an iteration of the same elementary map. Work on these
lines is in progress.

\bigskip
\bigskip

\noindent
{\Large{{\bf Acknowledgments}}}

\bigskip

\noindent
 A.B. has been
partially supported by Ministerio de Ciencia y Tecnolog\1a (Project
BFM2000-1055). O.R. has been partially supported by INFN and by MIUR
(COFIN2001 ``Geometry and Integrability"). 
Finally, we would like to thank Francisco J. Herranz for having
suggested
the  subalgebra method to construct coaction maps.

\end{document}